\documentclass[preprint,12pt,authoryear]{elsarticle}

\usepackage{amsmath,amssymb,amsfonts}
\usepackage{graphicx}
\usepackage{subcaption}
\usepackage{booktabs}
\usepackage{hyperref}
\usepackage{natbib}

\usepackage[T1]{fontenc}
\usepackage{titlesec}
\usepackage[a4paper, total={6in, 8in}]{geometry}

\usepackage{amssymb}
\usepackage{todonotes}
\usepackage{url}

\usepackage[utf8]{inputenc}
\usepackage[english]{babel}
\usepackage[T1]{fontenc}
\usepackage{fancyhdr}
\usepackage{lmodern}
\usepackage{multicol}
\usepackage{caption}
\usepackage{placeins}
\usepackage{array}
\usetikzlibrary{shapes,arrows}  

\usepackage{amsmath}

\usepackage{enumitem} 

\usepackage{makecell}
\usepackage{tabularx, ragged2e}

\let\cite\citep

\begin{document}

\begin{frontmatter}

\title{Sparse Sensor Placement in Multi-Agent Reinforcement Learning Control of Rayleigh-Bénard Convection}
\author[upb,lamarr]{Jan Stenner}

\author[upb,lamarr]{Hans Harder}

\author[tud,lamarr]{Sebastian Peitz}

\address[upb]{Universit\"{a}t Paderborn, Pohlweg 51, 33098 Paderborn, Germany}
\address[tud]{TU Dortmund, Joseph-von-Fraunhofer-Stra\ss{}e 25, 44227 Dortmund, Germany}
\address[lamarr]{Lamarr Institute for Machine Learning and Artificial Intelligence, Joseph-von-Fraunhofer-Stra\ss{}e 25, 44227 Dortmund, Germany}

\begin{abstract}
This paper studies sparse sensor placement for control of Rayleigh-Bénard convection with multi-agent reinforcement learning. We train dense expert policies with windowed observations and distill sparse apprentice policies by supervised learning with grouped regularization on encoder input weights. The framework combines ordered non-convex grouped regularization and iterative reweighted grouped regularization, and uses a grouping construction that enforces consistent pruning across overlapping observation windows. Experiments with fixed and varying initial conditions show that Multi-Agent Transformer policies train more stably than proximal policy optimization baselines, while sparse apprentices retain control behavior comparable to dense experts. Sparsity results are strong for the proposed grouped methods across settings, including maximal sparsity in all fixed-initial-condition setting variants and maximal or near-maximal sparsity in varying-initial-condition setting variants. As an additional proof of concept, training from learned minimal sensor sets reduces per-agent observation size from 360 to 12 and preserves the overall training trend in simulation while reducing data throughput. The results provide both an interpretable basis for identifying control-relevant spatial regions and state components, and a practical pathway toward sensor-efficient control under realistic hardware constraints.
\end{abstract}

\begin{keyword}
sparse sensor placement \sep partial differential equation control \sep Rayleigh-Bénard convection \sep multi-agent reinforcement learning \sep policy distillation \sep structured sparsification
\end{keyword}

\end{frontmatter}

\newpage

\section{Introduction}

Controlling systems governed by Partial Differential Equations (PDEs) is challenging because states are possibly high-dimensional, sensing is limited, and the dynamics are nonlinear and often chaotic.
This is particularly true for Rayleigh-B\'enard convection (RBC), a canonical buoyancy-driven flow where practical control is usually limited to changes in boundary temperatures.
Classical work on sparse PDE control and sparse sensor placement has shown that both actuation and sensing can be drastically reduced without losing essential dynamical information \cite{Casas2017,doi:10.1137/15M1036713,2018Manohar}.
These insights motivate data-driven control strategies that explicitly acknowledge partial observability.

Learning-based control has progressed from early machine-learning control ideas for nonlinear dynamics \cite{Duriez2017} to deep reinforcement learning (DRL) demonstrations on chaotic PDEs and flow control with limited sensors \cite{10.1098/rspa.2019.0351,Rabault_2019}.
The impact of sensor sparsity on performance and the need for memory/attention in partially observable regimes have been quantified recently \cite{Weissenbacher2025}, whereas our focus is on actively selecting sparse sensor subsets and distilling a compact multi-agent controller rather than only characterizing performance degradation.
For RBC specifically, multi-agent RL frameworks that exploit translational invariance enable scalable control with shared policies \cite{peitz2023}, yielding effective 2D RBC control \cite{vignon2023}, accelerated learning via positional encoding \cite{jeon2024}, and extensions toward turbulent regimes \cite{markmann2025controlrayleighbenardconvectioneffectiveness}.
These developments set the stage for structured, scalable control of RBC under severe sensing constraints.

At the same time, practical deployment calls for compact, interpretable policies and explicit sensor sparsification.
Sparse and structured neural-network regularization has been advanced through pruning and compression \cite{han2016deepcompressioncompressingdeep,obandoceron2024valuebaseddeepreinforcementlearning}, differentiable $L_0$ regularization \cite{louizos2018learningsparseneuralnetworks,botteghi2024parametricpdecontroldeep}, and group-structured parameter tying with GrOWL \cite{Oswal2016,Zhang2018,hotegni2024multiobjectiveoptimizationsparsedeep}.
Complementary to architectural sparsity, expert-apprentice paradigms in imitation learning and RL provide a mechanism to distill high-capacity teachers into smaller students \cite{pmlr-v15-ross11a,rusu2016policydistillation,berseth2018progressivereinforcementlearningdistillation,Livne_2020}.

In this work we focus on the control of RBC and combine these threads into a unified framework.
Building on the distributed multi-agent formulation of \cite{peitz2023} and the RBC setting of \cite{vignon2023}, we employ positional encoding \cite{jeon2024} and a Multi-Agent Transformer (MAT) architecture \cite{Wen2022} while introducing a teacher-student training pipeline that explicitly sparsifies sensors.
We further incorporate a small set of implementation changes to MAT that significantly stabilizes training in our RBC setting.
An expert policy is first trained with dense observations. Then a sparse apprentice is learned via supervised distillation with either ordered non-convex grouped regularization inspired by GrOWL \cite{Zhang2018, Oswal2016} or an iterative reweighted grouped regularizer \cite{candes2007enhancingsparsityreweightedl1} to select informative sensor subsets while pruning non-necessary ones.
We demonstrate effectiveness both for fixed initial conditions, as commonly assumed in RBC control studies \cite{vignon2023,jeon2024}, and for varying initial conditions across episodes, as has been done in newer works \cite{markmann2025controlrayleighbenardconvectioneffectiveness, becktepe2026plugandplaybenchmarkingreinforcementlearning}.
The resulting controller preserves performance while substantially reducing sensing and model complexity.

Beyond obtaining sparse controllers, the proposed framework provides two forms of practical value.
First, it provides a systematic framework for identifying the spatial regions and physical state components that are most informative for a given control objective in PDE-governed dynamical systems.
In our RBC example, these components are temperature and the two velocity components.
Second, it provides a pathway from simulation toward real deployments where dense sensing across the full domain is often infeasible or too costly. Our sparsification setup explicitly supports both hardware scenarios: systems with multi-component sensors that measure several state components at one location, and systems where each measured state component requires its own sensing unit and therefore contributes separately to installation effort.
The remainder of the paper is organized as follows: Section~\ref{sec:relatedwork} reviews the relevant foundations in RL, multi-agent PDE control, and sparsification. Section~\ref{sec:methodology} details the proposed training pipeline and architectures. Section~\ref{sec:experiments} presents the full experimental study, including the simulation setup, MAT-versus-PPO training behavior, expert-versus-apprentice trajectory comparisons for fixed and varying initial conditions, sparsity evaluations under multiple metrics, visualizations of the resulting sensor-placement patterns, and RL training under minimized sensor sets. Section~\ref{sec:conclusion} concludes.
Our code is publicly available at \href{https://github.com/janstenner/RBC-RL-SparseSensors}{https://github.com/janstenner/RBC-RL-SparseSensors}.

\section{Related Work}\label{sec:relatedwork}

In the following, we provide a brief overview of the key methods and concepts that form the foundation of our approach and are adapted or combined in the subsequent sections.

\subsection{Rayleigh-B\'enard Convection}\label{sec:rbc}

\begin{figure}[t]
	\centering
	\subcaptionbox{}{\includegraphics[width=0.32\columnwidth]{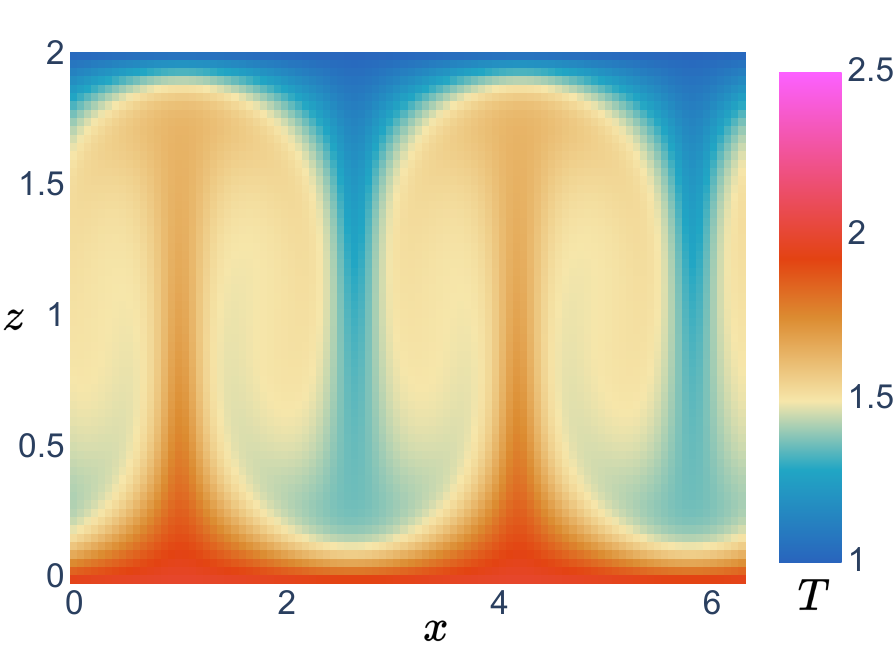}}
	\hfill
	\subcaptionbox{}{\includegraphics[width=0.32\columnwidth]{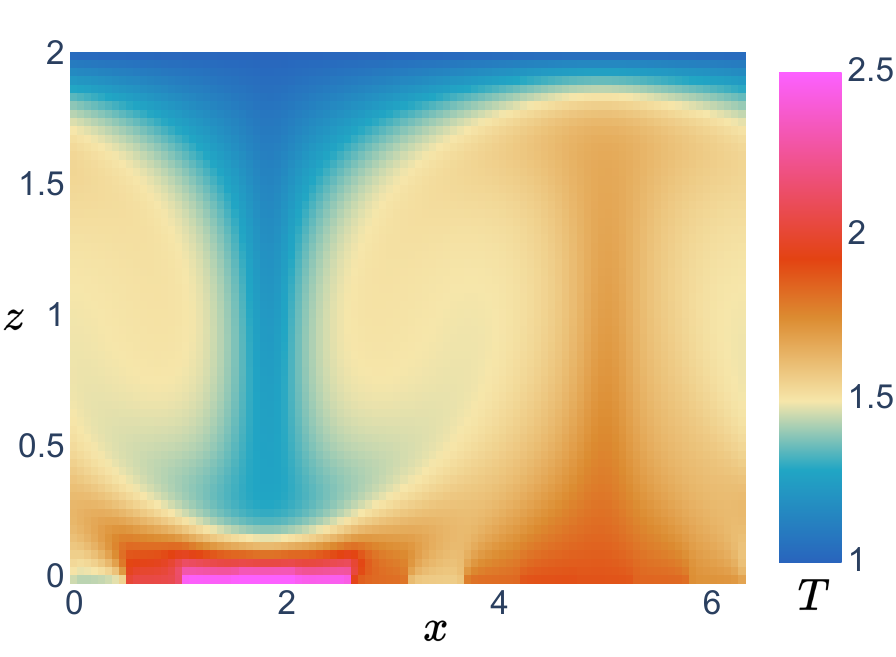}}
	\hfill
	\subcaptionbox{}{\includegraphics[width=0.32\columnwidth]{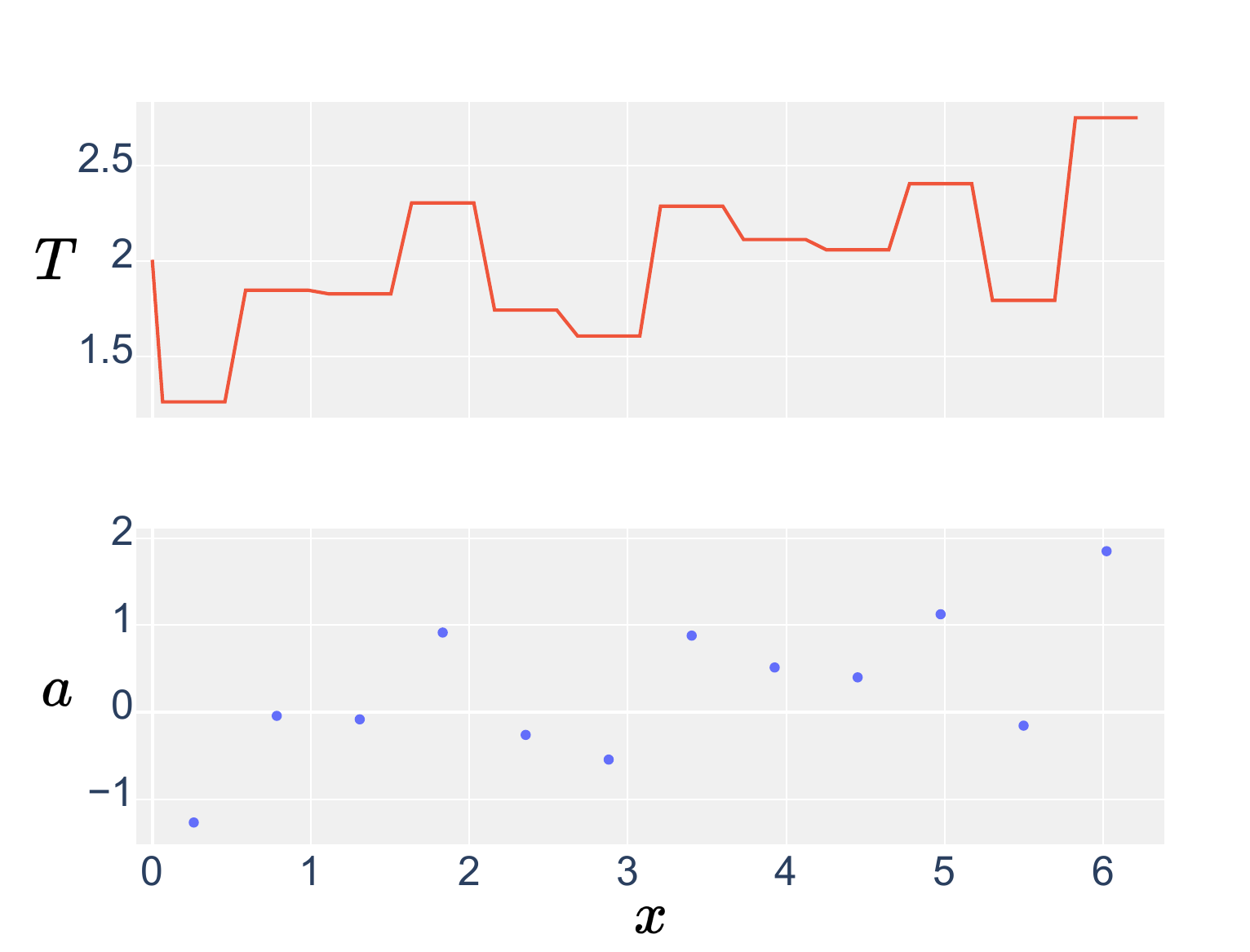}}
	\caption{Illustration for the Rayleigh-B\'enard convection control setup referenced in Section~\ref{sec:rbc}. The figure shows (a) the uncontrolled two-plume state of temperature, (b) a controlled one-plume state, and (c) an example of agent actions (bottom) together with the resulting boundary-control signal (top).}
	\label{fig:rbc_setup}
\end{figure}

Rayleigh-B\'enard convection (RBC) is a buoyancy-driven flow in which a fluid layer is heated from below and cooled from above, leading to formations of convection rolls over time. In the two-dimensional wide-aspect-ratio setups used in recent control studies, the lateral boundaries are typically periodic and the top/bottom walls are no-slip and isothermal. The uncontrolled flow at moderate Rayleigh numbers settles into a two-plume (double-cell) state, while feedback control can collapse the rolls into a single dominant plume and reduce convective heat transfer. Figure~\ref{fig:rbc_setup} provides a visual reference for these canonical states and the associated control signal.

Under the Oberbeck-Boussinesq approximation, the nondimensional 2D governing equations are
\begin{align}
	\nabla \cdot \mathbf{u} &= 0, \label{eq:rbc_continuity}\\
	\partial_t \mathbf{u} + (\mathbf{u}\cdot\nabla) \mathbf{u} &= -\nabla p + \sqrt{ \frac{ \mathrm{Pr} }{ \mathrm{Ra} } } \, \nabla^2 \mathbf{u} + T \, \mathbf{e}_2, \label{eq:rbc_momentum}\\
	\partial_t T + \mathbf{u}\cdot\nabla T &=  \frac{ 1 }{ \sqrt{ \mathrm{Ra} \, \mathrm{Pr} } } \, \nabla^2 T, \label{eq:rbc_energy}
\end{align}
where $\mathbf{u}=(u_1,u_2)$ is the velocity in the $(x_1,x_2)$ plane, $p$ the kinematic pressure, $T$ the nondimensional temperature (up to an arbitrary constant shift), and $\mathbf{e}_2$ the upward unit vector. We consider the rectangular domain $x_1\in[0,L_{x_1}]$ and $x_2\in[0,L_{x_2}]$, where $L_{x_1}$ and $L_{x_2}$ denote the domain length and height. The Rayleigh and Prandtl numbers are $\mathrm{Ra}=g\alpha \delta T H^3/(\nu \kappa)$ and $\mathrm{Pr}=\nu/\kappa$, with domain height $H$, temperature difference $\delta T$, kinematic viscosity $\nu$, and thermal diffusivity $\kappa$. We assume periodicity in $x_1$, no-slip walls at $x_2\in\{0,L_{x_2}\}$, a fixed top temperature $T_t$, and a controlled bottom temperature $T(x_1,0,t)=T_b + T_c(x_1,t)$ with $T_b-T_t=\delta T$. The control term is normalized over $x_1$ such that
\[
\int_0^{L_{x_1}} T_c(x_1,t)\,dx_1 = 0.
\]

The objective of control is to minimize the dimensionless heat transport measured by the instantaneous Nusselt number
\begin{equation}
	\mathrm{Nu}_\text{instant} = \sqrt{ \mathrm{Ra} \, \mathrm{Pr} } \left\langle u_2 T \right\rangle_{x_1,x_2}.
	\label{eq:rbc_nusselt}
\end{equation}
Here, $\langle\cdot\rangle_{x_1,x_2}$ denotes the spatial average over $x_1$ and $x_2$.

The sensor and actuator layout follows the standard benchmark configuration used in these works, in particular the setup described by Becktepe \cite{becktepe2026plugandplaybenchmarkingreinforcementlearning}. Control is applied through temperature modulation at the bottom boundary, which is discretized into equally spaced actuator segments along the horizontal direction. Each agent outputs a scalar action for its segment. These actions are mapped to a continuous boundary-control profile via interpolation or filtering before being imposed on the wall. Observations are drawn from a fixed, dense set of probe locations in the bulk that measure temperature and the two velocity components. The full sensor set forms a global grid of probes, and each agent receives a local window of contiguous probes centered on its actuator segment, consistent with the translationally invariant multi-agent design (\ref{sec:framework}).

\subsection{Reinforcement Learning}\label{sec:rl}

Here we will note the fundamentals of reinforcement learning (RL) that the remaining work of this paper is relying on. A thorough introduction with more explanations about concepts and methods can be found in \cite{SB18}. In reinforcement learning we typically have an \emph{agent} that is making sequential decisions at discrete time steps by interacting with an \emph{environment} in order to maximize cumulative reward. To do so, the agent receives a state $s_t$ at time step $t$ and after deciding on an action $a_t$ it receives a reward $r_{t+1}$ and transitions to the next state $s_{t+1}$. The goal of the agent is to find a policy $\pi(a|s)$, which is a probability distribution over $a$ given state $s$, that maps states to actions with the goal of maximizing the expected return, here defined as the sum of discounted future rewards.

Typically the dynamics of an environment are modeled as a Markov Decision Process (MDP)
\begin{align*}
    \big(  \mathcal{S}, \mathcal{A}, P, R, \gamma\big),
\end{align*}
where $\mathcal{S}$ is a (possibly finite) set of states, $\mathcal{A}$ is the set of actions available to the agent, $P(s'|s,a)$ is the state transition function giving the likelihood of transitioning to state $s'$ after taking action $a$ in state $s$, $R(s,a,s')$ is the reward function giving the reward after taking action $a$ in state $s$ and transitioning to $s'$, and $\gamma\in[0,1]$ is a discount factor.
The previously mentioned expected return is measured by the \emph{value function}, defined by $V^\pi(s_t) = \mathbb{E}_{\pi} \left[\sum_{n=t}^{\infty} \gamma^{n-t} R(\mathbf{s}_n,\mathbf{a}_n,\mathbf{s}_{n+1})\mid \mathbf{s}_t=s_t\right]$, where $\mathbf{s}_{n+1} \sim P(\cdot|\mathbf{s}_n,\mathbf{a}_n)$ and $\mathbf{a}_n \sim \pi(\cdot|\mathbf{s}_n)$.
Closely related to the value function is the \emph{Q function} or \emph{action-value function}, $Q^\pi(s_t,a_t)$, which estimates the expected return of taking action $a_t$ in state $s_t$ and then continuing under policy $\pi$.

In recent years, Deep Reinforcement Learning (DRL) has combined function approximation via deep neural networks with the exploration-vs-exploitation nature of RL. In this work we will use the Proximal Policy Optimization algorithm (PPO, see \cite{SchulmanWDRK17}), a DRL actor-critic method that represents both the policy and the value function with neural networks.

\subsection{Multi-Agent Framework for PDE Systems}\label{sec:framework}

\begin{figure}[t]
	\centering

    \subcaptionbox{}{\includegraphics[width=0.48\columnwidth]{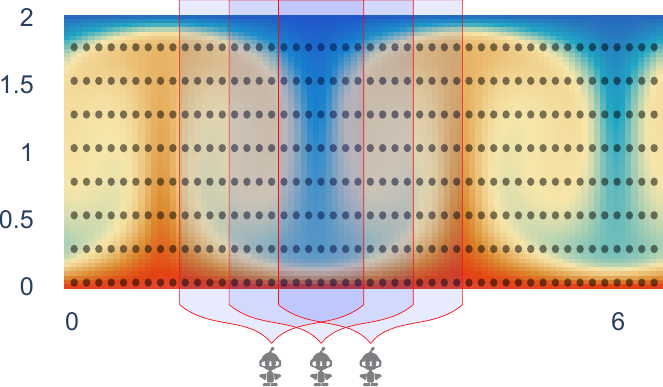}}
    \hfill
    \subcaptionbox{}{\includegraphics[width=0.48\columnwidth]{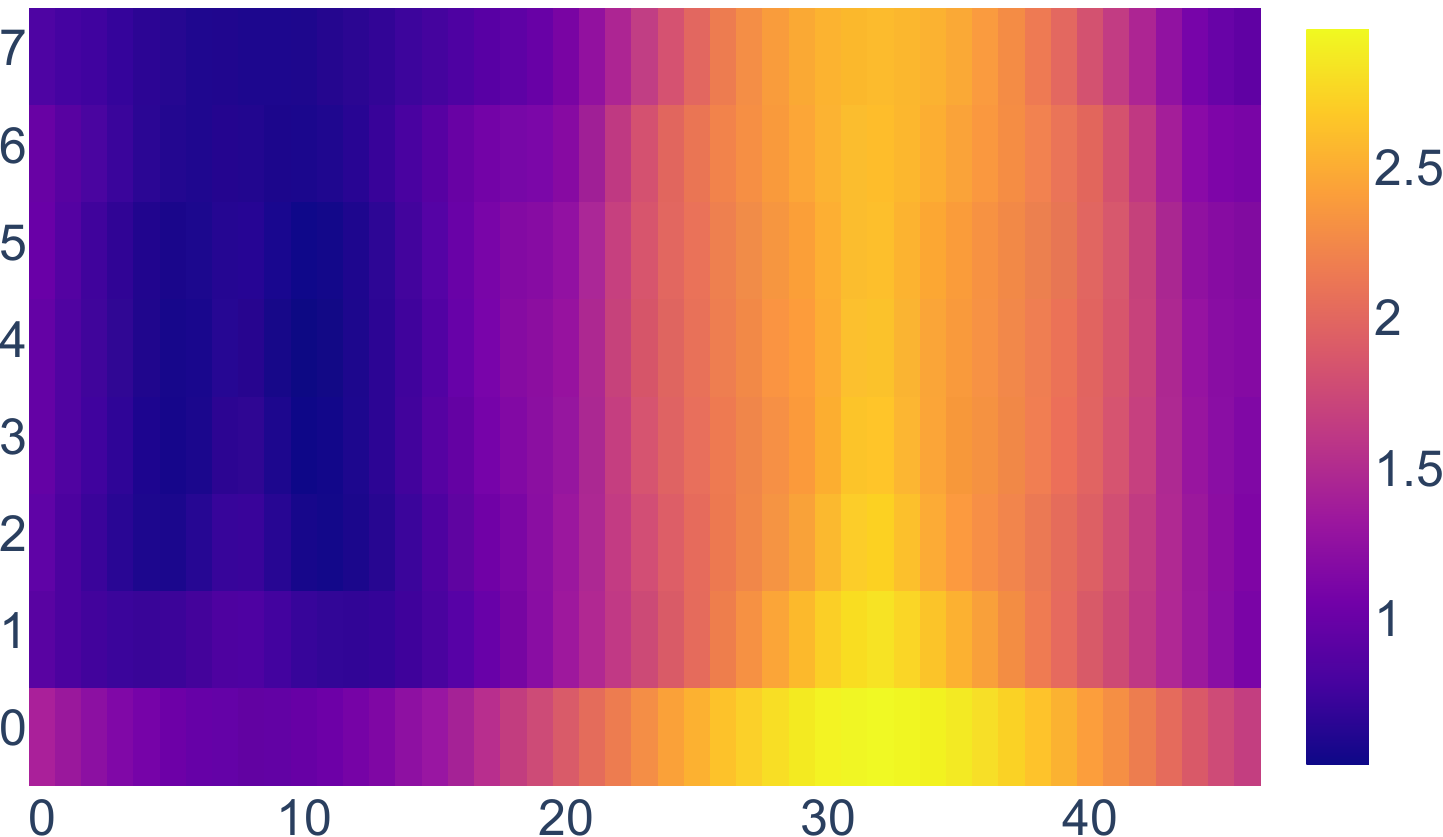}}
    \caption{(a) Schematic view of different agents and their observation windows of sensor sets for the Rayleigh-Bénard-Convection control problem. Only three agents are depicted for clarity. (b) Sensor‐window view of a temperature channel spanning 47 sensors with added positional encoding. Note that observations also include two velocity channels which are omitted here.}
  
	\label{fig:AgentWindows}
\end{figure}

The RL experiments in this paper follow the framework introduced in \cite{peitz2023} and \cite{vignon2023}. Agents observe the readings of sensors within a window of certain size, which is centered around the agents actuator positions, see figure \ref{fig:AgentWindows}.
Agents receive partial state sets, called observations, that are extracted according to their corresponding actuator positions.

To accurately model this framework we can extend the MDP with ideas from \emph{Partially Observable Markov Decision Processes} (POMDP, \cite{Spaan2012}) and \emph{stochastic games} (see \cite{Busoniu2010}) in the form of 
\begin{align*}
    \big(  \mathcal{S}, \mathcal{O}_1, \ldots , \mathcal{O}_{N_{\mathrm{ag}}}, \mathcal{A}_1, \ldots , \mathcal{A}_{N_{\mathrm{ag}}}, P, \Omega, R_1, \ldots , R_{N_{\mathrm{ag}}}, \gamma\big).
\end{align*}
Here, $\mathcal{O}_i$ are the sets of observations for each agent, $\mathcal{A}_i$ are the sets of actions, $R_i$ are the different reward functions, $\Omega: \Omega(s,i)=o_i$ is the (deterministic) observation function mapping the full state $s$ and agent index $i$ to an observation $o_i\in\mathcal{O}_i$, and ${N_{\mathrm{ag}}}$ is the number of agents. Since the agents in our framework have the same shape (and are, in fact, sharing parameters) we can assume that the sets $\mathcal{O}_i$ and $\mathcal{A}_i$ are the same for all $i$.

\subsection{Pruning and Sparsification}\label{sec:pruning}

To determine a sparse set of sensors that still enable us to have a controller that achieves performance comparable to that of the fully trained RL agent on a full sensor set, we train an apprentice model using supervised learning. We regularize groups of row vectors in a weight matrix $B$, which acts on a flattened input $x$ in the form of $x^T \mapsto x^T B$. Here, grouping is based on the magnitude of the row $\ell_2$ norms.
In our setup, $B$ is the weight matrix of the first layer of the neural network that parameterizes the actor policy.
Each row contains exactly the parameters that are multiplied with one input. Pruning a full row therefore removes one complete input from the model. Additionally, we group multiple rows together, so that structures tied to multiple inputs can be pruned jointly.
We consider two complementary regularization schemes, namely \emph{GrOWL-inspired Group-Ordered (GO)} and \emph{Iterative Grouped-Reweighted (GR)}.

For GO we build on the \emph{group ordered weighted $\ell_1$} (GrOWL, see \cite{Zhang2018, Oswal2016}) regularizer, which extends OWL (\cite{Figueiredo2016, zeng2015orderedweightedell1norm}). In our notation, the GO regularizer is
\begin{align*}
\mathcal{R}^{\mathrm{GO}}_\lambda(B)=\sum_{i=1}^{n}\lambda_i\lVert b_{[i]\cdot} \rVert_2,
\end{align*}
where $B\in\mathbb{R}^{n\times m}$, $\lambda \in \mathbb{R}^n$ with $0<\lambda_1\geq\lambda_2\geq\ldots\geq\lambda_n\geq0$ and $b_{[i]\cdot}$ is the row vector of $B$ with the $i$-th largest $\ell_2$ norm. Its behavior is influenced by the choice of the parameter vector $\lambda$ and can result in group-Lasso (or $\ell_{2,1}$) for $\lambda_1=\lambda_n$.
Intuitively, with an appropriate choice of $\lambda$, GrOWL applies stronger shrinkage to rows with larger norms while already small rows are penalized less. This can keep strongly correlated rows similar instead of pruning them independently as in standard group-Lasso.

While still following the GrOWL ordering principle on row-group norms, we choose a linearly increasing non-convex sequence. This design increases pressure on small-norm groups and pushes them closer to zero until they can be ignored in practice:
\begin{align}
0 \leq \lambda_1 \leq \lambda_2 \leq \ldots \leq \lambda_n, \notag \\
\lambda_i=\frac{i-1}{n}, \qquad i=1,\ldots,n.
\label{eq:lambda_linear}
\end{align}
Hence, the sequence runs from $\lambda_1=0$ to $\lambda_n=(n-1)/n$, which induces the non-convex weighting behavior in line with \cite{Huang2015}.

For GR we adopt the iterative reweighted $\ell_1$ method from \cite{candes2007enhancingsparsityreweightedl1} and transfer it to a grouped setting by replacing absolute values with group $\ell_2$ norms. In this grouped variant, groups are not ordered by magnitude.
Instead, we update the reweighting vector iteratively. For each iteration step $\ell$, we keep $\xi^{(\ell)}$ fixed in the regularizer, optimize $B$ for this weighted problem, and then compute $\xi^{(\ell+1)}$.
At step $\ell$ we use
\begin{align*}
\mathcal{R}^{\mathrm{GR}}_{\xi^{(\ell)}}(B)=\sum_{i=1}^{n}\xi_i^{(\ell)}\lVert b_{i\cdot}\rVert_2,
\end{align*}
with initial weights $\xi_i^{(1)}=1$ for all $i$ and the update rule
\begin{align}
\xi_i^{(\ell+1)}=\frac{1}{\lVert b_{i\cdot}^{(\ell)}\rVert_2+\varepsilon}, \qquad \varepsilon>0.
\label{eq:reweight_update}
\end{align}
Here, $b_{i\cdot}$ denotes the (non-ordered) $i$-th row of $B$. Therefore, large groups are downweighted and small groups are penalized more strongly over iterations, promoting sparsity without using any ordering of groups.

\subsection{Multi-Agent Transformer}\label{sec:mat}

\begin{figure}[t]
	\centering

    \includegraphics[width=0.7\columnwidth]{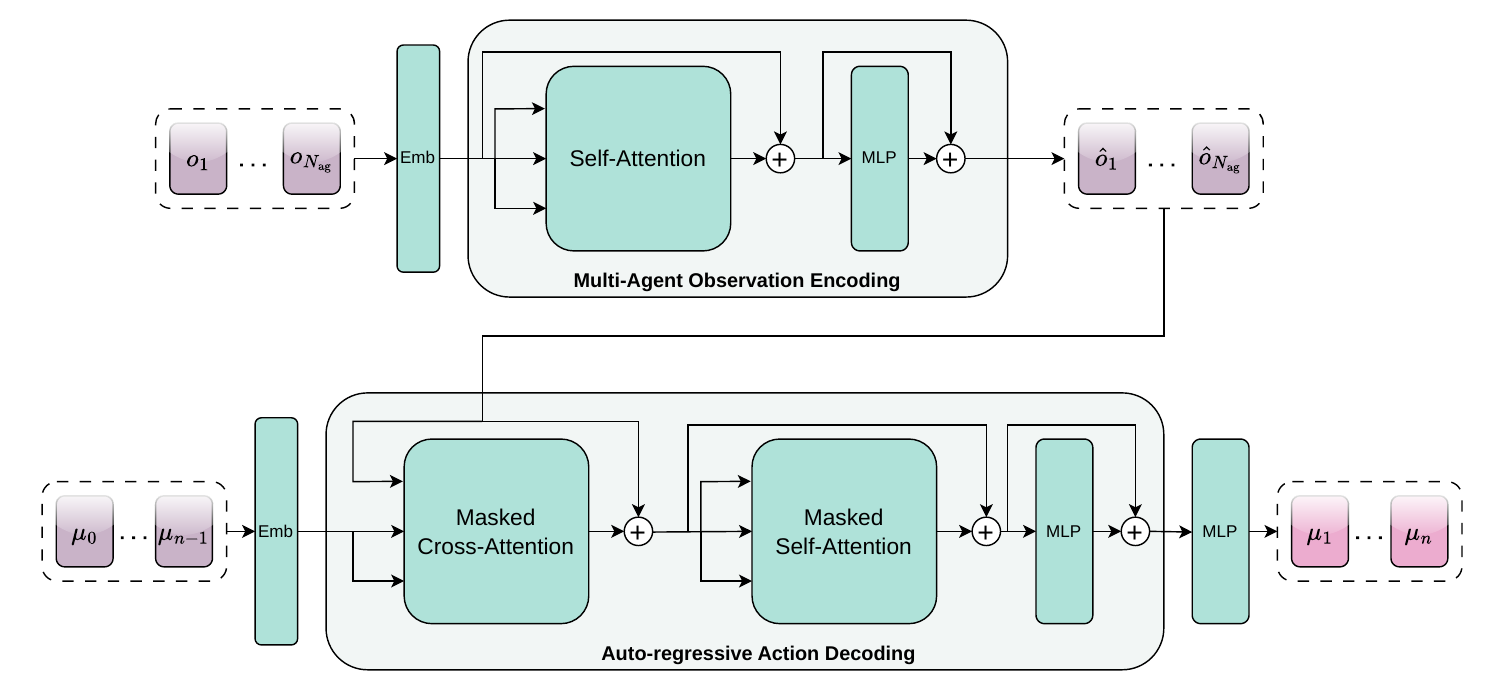}
    \caption{Architecture of the Multi-Agent Transformer (MAT) network used as an apprentice model for supervised learning from an expert. We illustrate this with the generation of action tokens $\mu_1$ to $\mu_n$ in the decoder, which are the means of the output normal distributions used to sample actions. In the auto-regressive framework, this effectively amounts to generating $\mu_n$, as earlier tokens are treated as fixed context. Note that $\mu_0$ is the start token that we choose to be $[0.0]$.}
  
	\label{fig:CustomMat}
\end{figure}

In this work, we use a slightly modified \emph{Multi-Agent Transformer} (MAT, see \cite{Wen2022}) in two roles: (i) as a policy architecture trained directly with PPO and compared against direct PPO training, and (ii) as a candidate expert policy for later distillation. The apprentice architecture is depicted in figure \ref{fig:CustomMat}. While being labelled a Multi-Agent method, it bridges the gap between Multi-Agent and Single-Agent setups due to information sharing between agents. We include MAT as an expert candidate because in our varying-initial-condition setting it is empirically more stable than plain PPO.

In MAT, the observations of all agents are first processed jointly by the encoder.
Given the observation sequence $\mathbf o = (o_1,\ldots,o_{N_{\mathrm{Ag}}})$, the encoder produces latent representations $\hat{\mathbf o}$, where each $\hat{o}_i$ contains a contextualized, latent representation of agent $i$'s observation.
The original MAT draws one action per agent
\begin{align}
    a_i \sim \pi_i(\cdot | \hat{\mathbf o}, \mathbf a_{1:i-1}),
    \label{eq:mat_action_distribution}
\end{align}
where $\mathbf a_{1:i-1}$ is the sequence of actions of all previously evaluated agents, without prescribing any fixed order of agents.
The distribution in \ref{eq:mat_action_distribution} is the output of the decoder, which takes the encoded observation representations $\hat{\mathbf o}$ and the previously drawn actions $\mathbf a_{1:i-1}$ as input.
After drawing action $a_i$, it is prepended to the sequence of previously drawn actions, i.e. drawing from distributions is part of the autoregressive sequence generation loop.

In our implementation, actions are not immediatly drawn after individual distributions are generated.
Instead, the decoder generates the next distribution based on the previously generated distribution sequence, and actions are drawn jointly from a multi-variate distribution after the sequence is completed.
Additionally, we keep a fixed order of agents.
When modeling this with a Gaussian distribution and keeping the standard derivation a learnable but global parameter, the decoder effectively generates the mean of the distribution for each agent.
Starting from an initial token $\mu_0$, the decoder then predicts the shifted sequence
\begin{align*}
\mu_0
&\longrightarrow
\mu_1, \\
\mu_0,\mu_1
&\longrightarrow
\mu_1,\mu_2, \\
\mu_0,\mu_1,\mu_2
&\longrightarrow
\mu_1,\mu_2,\mu_3, \\
&\ \ \vdots \\
\mu_0,\mu_1,\ldots,\mu_{N_{\mathrm{Ag}}-1}
&\longrightarrow
\mu_1,\mu_2,\ldots,\mu_{N_{\mathrm{Ag}}}.
\end{align*}
These modifications to the original MAT implementation have been found to be slightly more stable and allow for a more straightforward way of estimating deterministic joint expert actions in subsection \ref{sec:growlreg}.

We also introduce three more modifications to the original architecture:
First, we use a dedicated encoder branch for value prediction, in addition to the encoder that provides observation representations $\hat{o}_i$ to the decoder.
Second, we place cross-attention before self-attention and keep the MAT-specific cross-attention wiring
\begin{align*}
  \mathrm{Attention}(Q, K, V)
= \mathrm{softmax}\!\biggl(\frac{QK^\top}{\sqrt{d_k}}\biggr)\,V,
\end{align*}
which is the standard notation for attention (see \cite{vaswani2023}), where $d_k$ is the key dimension and the query matrix $Q$ is computed from the encoder's final observation representations while the key and value matrices $K$ and $V$ are computed from the decoder's layer input.
The observation representations are also used as the residual information that gets propagated via the skip-connection.
Third, we disable dropout and LayerNorm throughout the architecture, which further improves training behavior.

\section{Methodology}\label{sec:methodology}

We now proceed by introducing the three core methods used to obtain sparse sensor sets, along with controllers capable of operating on them while still achieving comparable performance. These are namely \emph{Reducing Window Size}, \emph{Apprentice Training with Group Regularization} and \emph{Grouping for Overlapping Windows}, the latter being an extension of the second method.

\subsection{Reducing Window Size}\label{sec:windowsize}

This step involves training of several RL agents with different window sizes. Due to the overlapping nature of windows in our framework, as described in \ref{sec:framework}, this only reduces sensors needed in the whole domain if windows can be reduced to a size in which overlapping does not occur. As documented in section~\ref{sec:experiments}, for the experiments presented in this work, this method alone does not achieve a reduced sensor set. It does, however, yield multiple agents that can take on the roles as experts in the next steps. Since it also serves as a form of investigatory analysis of the problem at hand, and could already provide sparser solutions for other experiments, we still consider it a valuable first step method.

\subsection{Apprentice Training with Group Regularization}\label{sec:growlreg}

For vector spaces $O$ and $A$, let $\pi_e:O\rightarrow A$ be a policy, which we call the \emph{expert}, and $f_\theta:O\rightarrow A$ a model parameterized by $\theta$, which we call the \emph{apprentice}.
We now formalize an optimization problem
\begin{align}
\min_\theta \sum_{i=1}^m \left\lVert f_\theta(o^{(i)}) - \pi_e(o^{(i)}) \right\rVert_2^2 + \mathcal{R}^{\mathrm{M}}(B), \qquad \mathrm{M}\in\{\mathrm{GO},\mathrm{GR}\}
\label{optiproblem}
\end{align}
for a data set $o^{(i)}$, $i=1,\ldots,m$.
Here, $\mathcal{R}^{\mathrm{GO}}=\mathcal{R}^{\mathrm{GO}}_\lambda$ and $\mathcal{R}^{\mathrm{GR}}=\mathcal{R}^{\mathrm{GR}}_{\xi^{(\ell)}}$ from Section~\ref{sec:pruning}.
For the GO case, rows are ordered by group norm through $b_{[i]\cdot}$.
For the GR case, rows are indexed directly by $b_{i\cdot}$ without ordering.
For the iterative reweighted case, we keep $\xi^{(\ell)}$ fixed for a certain number of updates $K_{\mathrm{prox}}$ within the same outer iteration level $\ell$.
After these $K_{\mathrm{prox}}$ inner updates, we update the weights according to (\ref{eq:reweight_update}).
$B$ denotes the first‐layer weight matrix of $f_\theta$, whose rows index the input neurons, and whose entries are part of the parameter vector $\theta$.
We solve (\ref{optiproblem}) with a proximal-operator algorithm, which alternates a gradient step for the prediction loss with a proximal shrinkage step that enforces the structured regularization and drives weak row groups toward zero.
Our implementation is based on the authors' public implementation\footnote{https://github.com/Dejiao2018/GrOWL}.

Since we use MAT as the apprentice architecture, which follows a sequence-based centralized-execution principle, the inputs $o^{(i)}$ consist of ordered sequences of local observations of all present agents for one decision step.
After successful completion of training, the shared actor parameters $\theta_e$ define the expert policy $\pi_{\theta_e}=\pi_e$ that serves as the common expert instance for all agents.
In our setting, $\pi_e(o^{(i)})$ denotes the sequence of mean values $\mu^{(i)}=[\mu^{(i)}_1,\ldots,\mu^{(i)}_{N_{\mathrm{ag}}}]^\top$, obtained by autoregressively processing $o^{(i)}$, which we use as a deterministic approximation of the expert policy.
During supervised training, the autoregressive decoder is evaluated with the additional shifted regression input $[\mu_0, \mu^{(i)}_1,\ldots,\mu^{(i)}_{N_{\mathrm{ag}}-1}]^\top$, where $\mu_0$ is the sequence starting token.
The resulting predicted sequence is compared component-wise against the expert target sequence $\mu^{(i)}$ in the mean-squared error term of (\ref{optiproblem}).

\subsection{Grouping for Overlapping Windows}\label{sec:groupingwindows}

\begin{figure}[t]
	\centering

    \includegraphics[width=0.7\columnwidth]{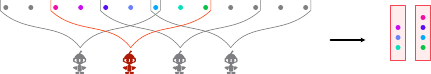}

    \caption{Overlapping windows with sensors in one dimension. The window of one agent is highlighted, and the sensors contained in this window are color-coded. The figure also shows how two grouping sets, indicated by red backgrounds, are constructed from these sensors. These sets are disjoint and together contain the local indices of all sensors in the highlighted window.}
  
	\label{fig:OverlappingProblem}
\end{figure}

Figure \ref{fig:OverlappingProblem} illustrates why naive row-wise pruning is insufficient in overlapping-window settings. The same physical sensor can appear under different local indices in different agent windows. If these indices are pruned independently, the corresponding sensor location may still remain active somewhere else.

To enforce consistent pruning, we construct grouping sets $\mathcal I_h$ consisting of indices of flattened inputs, which are also row indices of $B$, such that indices in the same class correspond to the same global sensor position under shifted windows. With correct construction these sets form equivalence classes that induce a relation
\begin{align*}
i \sim j \quad\Longleftrightarrow\quad \exists\,h:\; i,j\in\mathcal I_h,
\end{align*}
and partition the local index set into groups that should be pruned jointly. The full construction of these classes, including the general $N$-dimensional notation and the overlap proof, is provided in Appendix~\ref{app:grouping}.

We now extend the optimization problem in (\ref{optiproblem}) to grouped pruning.
Let $N_{\mathcal H}$ denote the number of non-empty grouping classes $\mathcal I_h$, and index these classes by $g\in\{1,\dots,N_{\mathcal H}\}$.
For each such class, let $\hat b_g$ denote the vector obtained by concatenating all row vectors $b_{i\cdot}$ with $i$ in that class, in any fixed order.
We then formulate the grouped optimization problem

\begin{align}
\min_\theta \sum_{i=1}^m \left\lVert f_\theta(o^{(i)}) - \pi_e(o^{(i)}) \right\rVert_2^2 + \hat{\mathcal R}^{\mathrm{M}}(B) &,\qquad \mathrm{M}\in\{\mathrm{GO},\mathrm{GR}\},
\label{optiproblem2}\\
\hat{\mathcal R}^{\mathrm{GO}}_\lambda(B) = \sum_{g=1}^{N_{\mathcal H}}\lambda_g\lVert \hat b_{[g]} \rVert_2 &, \notag\\
\hat{\mathcal R}^{\mathrm{GR}}_{\xi^{(\ell)}}(B) = \sum_{g=1}^{N_{\mathcal H}}\xi_g^{(\ell)}\lVert \hat b_{g} \rVert_2 &, \notag
\end{align}
where $\hat b_{[g]}$ denotes the $\hat b_{g}$ with the $g$-th largest $\ell_2$ norm.

\subsubsection{Input Channels}\label{sec:inputchannels}

One aspect that is especially relevant for dynamical systems governed by PDEs is that each sensor location can carry multiple measured state components. In our experiments, each location provides three components: temperature, horizontal velocity, and vertical velocity. In the default grouping used here, these components are treated as separate input entries at each location, so pruning can remove one component while keeping the others at the same spatial point.

In real scenarios it might, however, be of interest to group channels together at the same spatial locations. This yields two cases: channel-separated grouping, where channels are pruned independently, and channel-coupled grouping, where channel indices at one spatial location are pruned jointly. We use both cases in our experiments. The formal channel-coupled construction is given in Appendix~\ref{app:channel-grouping}.

\section{Experiments}\label{sec:experiments}

In this section, we describe the numerical setup used for all reported experiments. We run the simulations with Oceananigans.jl \cite{silvestri2024oceananigans} on a grid of size $96\times 64$ in a domain with $L_{x_1}=2\pi$ and $L_{x_2}=2$. The physical parameters are set to $\delta T = 1$, $\mathrm{Ra}=10^4$ and $\mathrm{Pr}=0.7$. We use a sensor grid of $48\times 8$ and $12$ actuators/agents, corresponding to one actuator per four sensor points in $x_1$ direction.

To initialize both reported experiment types, we first simulate $300$ non-dimensional time steps to obtain the canonical two-plume state. Each RL episode then spans $300$ non-dimensional time steps, and each agent applies one action every $1.5$ time steps, yielding $200$ control steps per episode. Rewards are defined as the negative instantaneous Nusselt number, $r_t=-\mathrm{Nu}_{\text{instant},t}$, with $\mathrm{Nu}_{\text{instant}}$ defined in Eq.~(\ref{eq:rbc_nusselt}) and evaluated from sensor measurements.
We apply positional encoding on the observations, which was shown to improve performance (see \cite{jeon2024}), allowing easier orchestration of asynchronous actions although agents share parameters. Since position information is already part of the observations, we do not add a separate positional encoding inside MAT.
Figure \ref{fig:AgentWindows} conceptually depicts how observations are extracted from the global sensor set and how the added positional encoding appears. Note that only the temperature channel is shown in that figure, while the experiments use two additional velocity channels. In the following, all experiments use a window size of $15$.

\begin{figure}[!htbp]
    \centering
    \includegraphics[width=0.8\columnwidth]{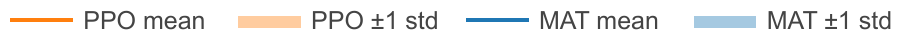}

    \subcaptionbox{}{%
        \includegraphics[width=0.48\textwidth]{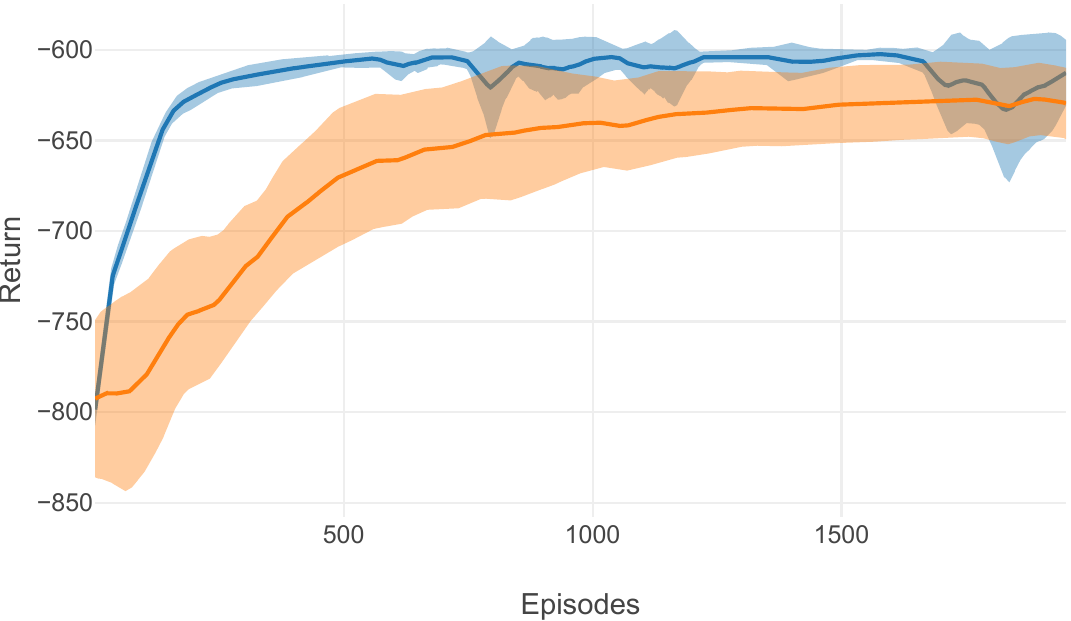}
    }
    \hfill
    \subcaptionbox{}{%
        \includegraphics[width=0.48\textwidth]{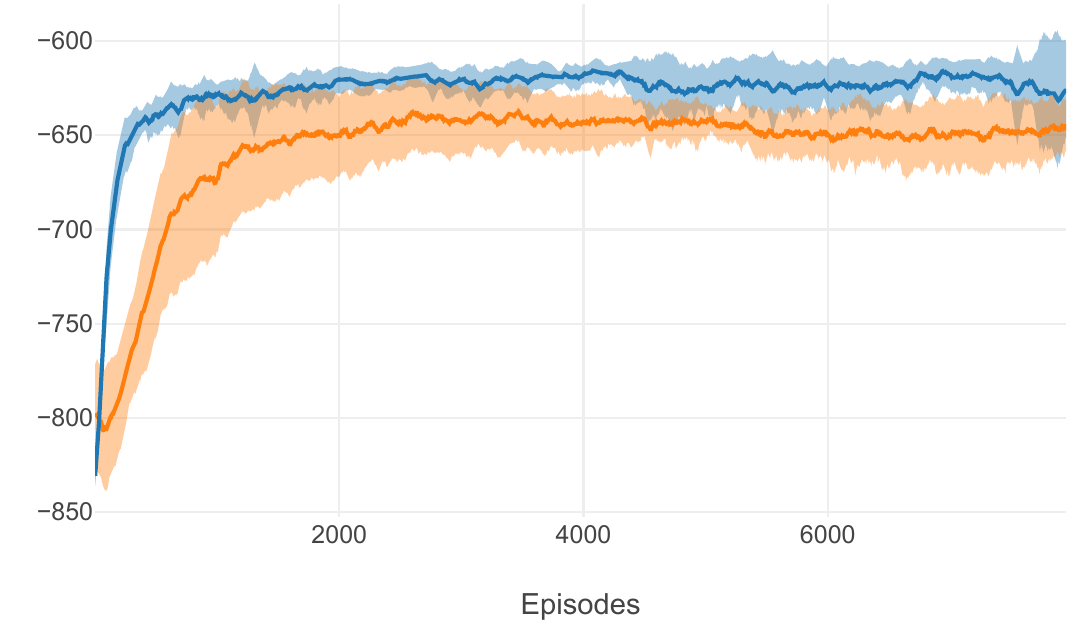}
    }
    \caption{MAT-versus-PPO training curves for both initial-condition settings. Panel (a) shows fixed initial conditions with mean and one-standard-deviation bands over 10 runs per algorithm, 2000 episodes per run, and smoothing over 50 measurements. Panel (b) shows varying initial conditions with the same statistics over 10 runs, 8000 episodes per run, and smoothing over 50 measurements. In both cases, MAT trains more stably and attains better performance than PPO.}
    \label{fig:mat_vs_ppo_both_ic}
\end{figure}

After expert RL training, we instantiate a separate MAT policy (encoder and decoder) as an apprentice, with configuration matched to the corresponding initial condition (IC) setting.
Training data for the apprentice is generated from expert rollouts.
For each expert action distribution, we denote by $\mu_{\mathrm{expert}}$ the mean of the output normal distribution from which the action is sampled.
Using these means as targets implements the deterministic policy evaluation used in the distillation objective.
The apprentice is trained by supervised learning with mean-squared error on predicted actions as objective.

In parallel, we apply structured pruning to encoder input weights using either group-ordered or group-reweighted regularization, with grouping defined by Section~\ref{sec:groupingwindows}. After each supervised backpropagation step, we apply proximal regularization steps to these grouped encoder-input rows. For the ordered case we use the non-convex sequence in Eq.~(\ref{eq:lambda_linear}). For the reweighted case we use Eq.~(\ref{eq:reweight_update}). In both cases, regularization weights in the proximal step are scaled by a configuration-dependent regularization power parameter, chosen strictly below $0.1$. Throughout training we monitor imitation loss and sparsity. After training, we generate a binary mask for the encoder input and keep the pair of apprentice policy and mask as the final artifact.

\subsection{Apprentice Distillation and Grouped Pruning Results}\label{sec:fixedIC}\label{sec:varyingIC}

We evaluate apprentice training in two initial-condition settings. For fixed initial conditions, training data is taken from one directly generated expert rollout episode. We collect data from 200 control steps and use a batch size of 20. The apprentice architecture is a compact MAT with an latent dimension of 32 and one transformer block. For varying initial conditions, training data is collected as a cache of 4000 control steps from 20 expert rollouts with random initial conditions, and we train with a batch size of 100. The apprentice uses a slightly larger MAT with latent dimension of 44.

\begin{figure}[!htbp]
    \centering

    \includegraphics[width=0.49\columnwidth]{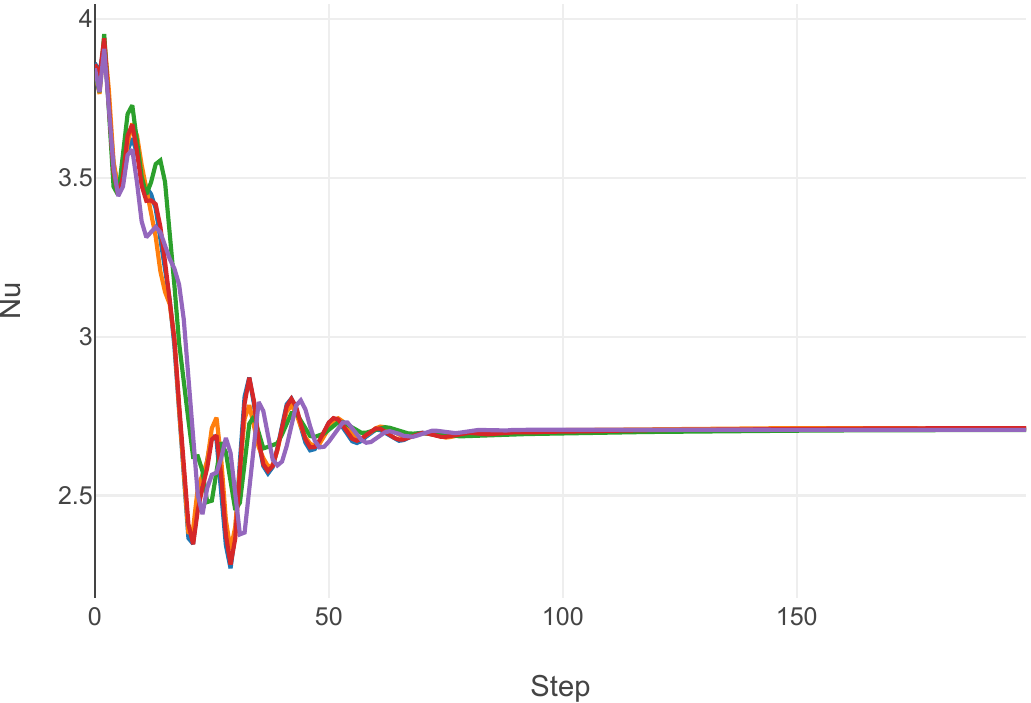}
    
    \includegraphics[width=0.8\columnwidth]{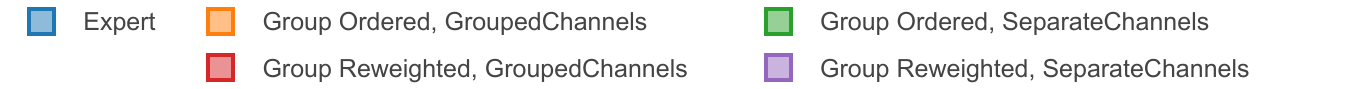}

    \caption{Global Nusselt-number trajectories over time for the fixed-initial-condition scenario, comparing the expert to all apprentice combinations.}
    \label{fig:fixed_ic_same_day_comparison}
\end{figure}

Figure~\ref{fig:mat_vs_ppo_both_ic} shows the RL training behavior (window size $15$) of the expert candidates in both settings and indicates that MAT is consistently more stable than PPO. This is consistent with the information exchange across observation tokens via attention described in Section~\ref{sec:mat}.

\begin{figure}[!htbp]
    \centering
    \includegraphics[width=0.8\columnwidth]{comparison_legend.pdf}

    \begin{minipage}[t]{0.49\textwidth}
        \vspace{0pt}
        \centering
        \subcaptionbox{}{%
            \begin{minipage}[c][0.30\textheight][c]{\linewidth}
                \centering
                \includegraphics[width=\linewidth,height=0.28\textheight,keepaspectratio]{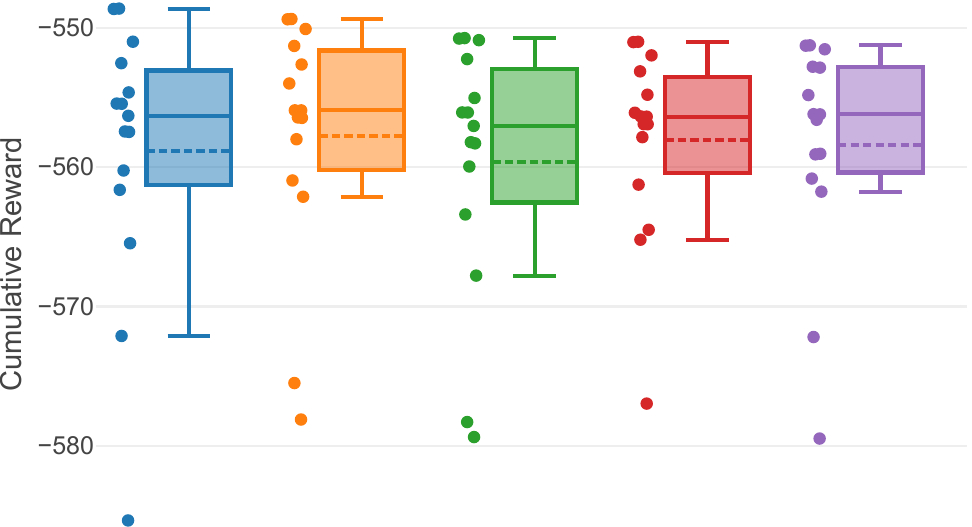}
            \end{minipage}%
        }
    \end{minipage}
    \hfill
    \begin{minipage}[t]{0.49\textwidth}
        \vspace{0pt}
        \centering
        \subcaptionbox{}{%
            \begin{minipage}[c][0.30\textheight][c]{\linewidth}
                \centering
                \includegraphics[width=\linewidth,height=0.28\textheight,keepaspectratio]{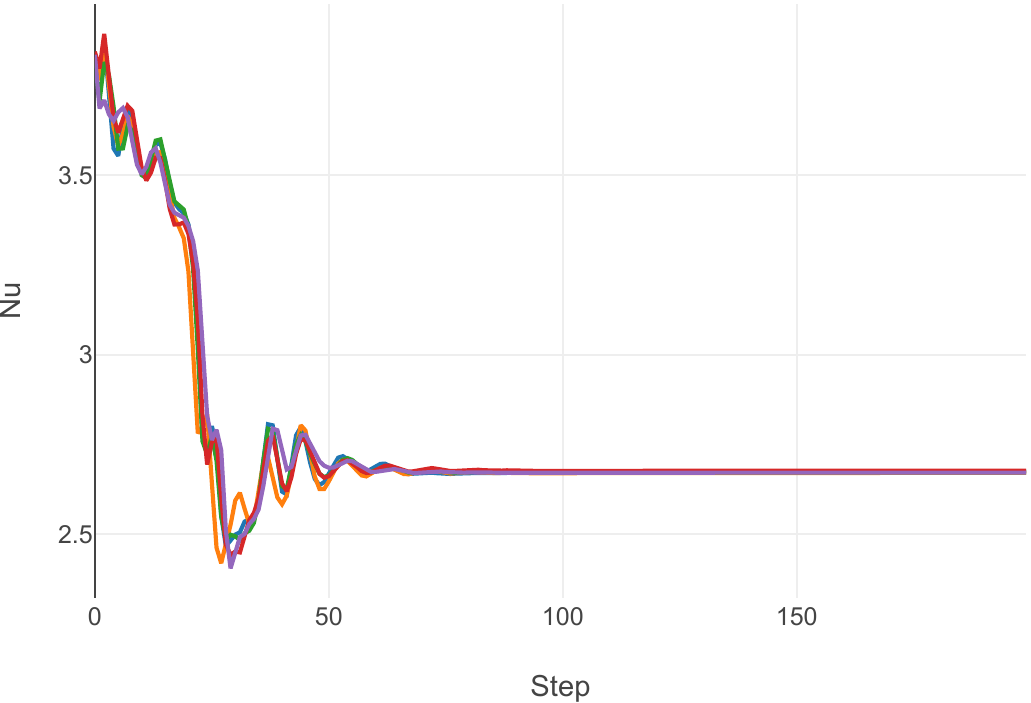}
            \end{minipage}%
        }
    \end{minipage}
    \caption{Varying-IC performance comparison. Panel (a) shows cumulative-reward box plots on the 15 fixed test initial conditions (offsets). Panel (b) shows the global Nusselt-number trajectory for one example initial condition from this set.}
    \label{fig:varying_ic_boxes_and_same_day}
\end{figure}

The resulting apprentice policies retain expert-like control behavior in both settings. Figure~\ref{fig:fixed_ic_same_day_comparison} compares global Nusselt-number trajectories for fixed initial conditions and shows that the trained apprentices control the system on a level comparable to the expert. Figure~\ref{fig:varying_ic_boxes_and_same_day} extends this comparison to varying initial conditions.
The figure reports cumulative rewards on a test set of $15$ fixed random offsets and includes one representative global-Nusselt trajectory from that set, showing that apprentice performance remains comparable to the expert.

Table~\ref{tab:sparsity_results} reports sparsity with three metrics:
\begin{itemize}
    \item \textbf{WS (Window Sparsity)} treats the three channels separately within each local window.
    \item \textbf{WS-CC (Window Sparsity combined channels)} overlays channels before computing sparsity in the local window.
    \item \textbf{TS-CC (Total Sparsity combined channels)} overlays channels on the global grid.
\end{itemize}
This distinction matters for deployment. With multi-sensors that can measure all channels at one location, combined-channel metrics are often the relevant criterion. Without such sensors, channel-separated sparsity can be operationally more important.

\begin{table}[!htbp]
    \centering
    \subcaptionbox{Fixed IC\label{tab:fixed_ic_sparsity}}[0.48\textwidth]{%
        \centering
        \scriptsize
        \resizebox{\linewidth}{!}{%
            \begin{tabular}{lccc}
                \hline
                Setting & WS (\%) & WS-CC (\%) & TS-CC (\%) \\
                \hline
                GO + SC & \textbf{98.889} & \textbf{96.667} & \textbf{96.875} \\
                GO + GC & 96.667 & \textbf{96.667} & \textbf{96.875} \\
                GR + SC & \textbf{98.889} & \textbf{96.667} & \textbf{96.875} \\
                GR + GC & 96.667 & \textbf{96.667} & \textbf{96.875} \\
                Lasso + SC & 81.667 & 47.500 & 46.875 \\
                Lasso + GC & 81.389 & 80.000 & 81.250 \\
                GrOWL + SC & 20.833 & 0.000 & 0.000 \\
                GrOWL + GC & 12.222 & 0.000 & 0.000 \\
                \hline
            \end{tabular}%
        }
    }
    \hfill
    \subcaptionbox{Varying IC\label{tab:varying_ic_sparsity}}[0.48\textwidth]{%
        \centering
        \scriptsize
        \resizebox{\linewidth}{!}{%
            \begin{tabular}{lccc}
                \hline
                Setting & WS (\%) & WS-CC (\%) & TS-CC (\%) \\
                \hline
                GO + SC & \textbf{97.778} & 93.333 & 93.750 \\
                GO + GC & 96.667 & \textbf{96.667} & \textbf{96.875} \\
                GR + SC & 96.667 & 90.000 & 90.625 \\
                GR + GC & 96.667 & \textbf{96.667} & \textbf{96.875} \\
                Lasso + SC & 69.722 & 32.500 & 34.375 \\
                Lasso + GC & 72.778 & 71.667 & 71.875 \\
                GrOWL + SC & 19.167 & 2.500 & 0.000 \\
                GrOWL + GC & 20.000 & 0.000 & 0.000 \\
                \hline
            \end{tabular}%
        }
    }
    \caption{Abbreviations: GO = Group Ordered, GR = Group Reweighted, SC = Separate Channels, GC = Grouped Channels, WS = Window Sparsity, WS-CC = Window Sparsity combined channels, TS-CC = Total Sparsity combined channels. Bold values mark the best value in each column.}
    \label{tab:sparsity_results}
\end{table}

The table also includes two comparison baselines. The Lasso baseline corresponds to the equal-$\lambda$ case described in Section~\ref{sec:pruning}. The GrOWL baseline uses the same $\lambda$ sequence as GO, but in descending order. Compared with GO and GR, Lasso reaches lower sparsity and GrOWL reaches substantially lower sparsity in both initial-condition settings.
The GO and GR method/channel-grouping combinations reach the maximal sparsity level attainable in our fixed-IC setup. In the varying-IC setting, GO and GR grouped-channel variants again reach maximal sparsity, while their separate-channel variants remain near-maximal. Within separate channels, group-ordered yields higher sparsity than group-reweighted.

For Lasso and GrOWL, we additionally set inputs with group norms close to zero exactly to zero after training. This post-training thresholding is required to obtain the sparsity values reported in Table~\ref{tab:sparsity_results}.
The threshold is chosen as high as possible while still retaining competitive control scores. This also explains why GrOWL with descending $\lambda$ values reaches $0\%$ for grouped channels in WS-CC and TS-CC.
Across the GrOWL runs, training itself does not drive inputs exactly to zero, and the thresholding operation is independent of the channel grouping. In contrast, GO and GR achieve the reported sparsity directly without a norm threshold.
In control regimes that are more unstable than the present $\mathrm{Ra}=10^4$ RBC benchmark, the less aggressive pruning behavior of Lasso and GrOWL may be useful because such methods can preserve control performance more reliably under harder dynamics.

\begin{figure}[!htbp]
    \centering
    \begin{minipage}[t]{\textwidth}
        \vspace{0pt}
        \centering
        \subcaptionbox{GO + GC\label{fig:fixed_sparse_go_gc}}{\includegraphics[width=0.48\linewidth]{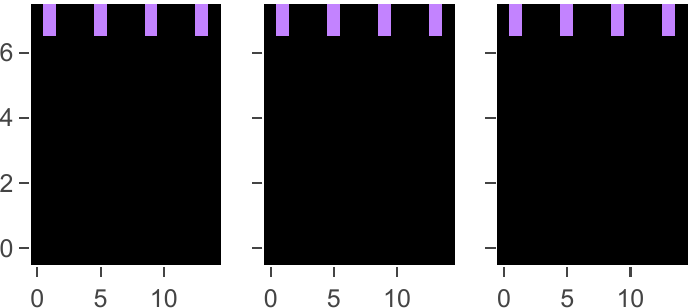}}
        \hfill
        \subcaptionbox{GO + SC\label{fig:fixed_sparse_go_sc}}{\includegraphics[width=0.48\linewidth]{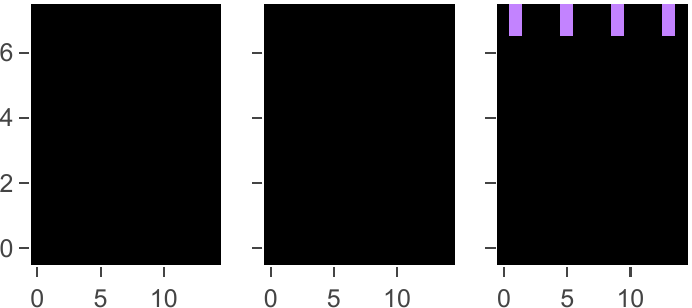}}

        \vspace{0.6em}

        \subcaptionbox{GR + GC\label{fig:fixed_sparse_gr_gc}}{\includegraphics[width=0.48\linewidth]{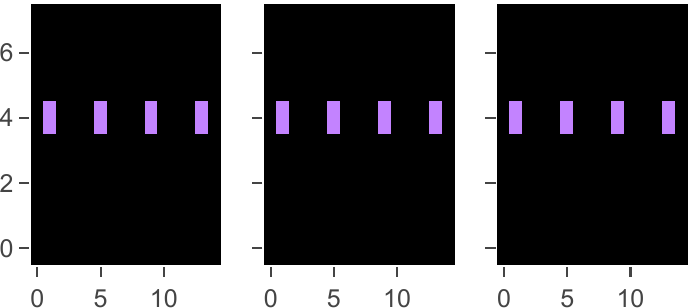}}
        \hfill
        \subcaptionbox{GR + SC\label{fig:fixed_sparse_gr_sc}}{\includegraphics[width=0.48\linewidth]{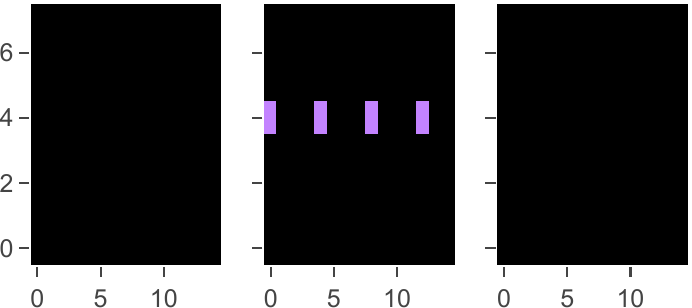}}
    \end{minipage}
    \hfill
    \caption{Fixed-IC window sparsity plots for the GO/GR regularizer and channel-grouping combinations. For each combination, three channels are shown in this order: temperature, vertical velocity, and horizontal velocity. The width of each observation plot corresponds to window size 15. Black cells mark pruned inputs, and light-violet cells mark non-pruned inputs.}
    \label{fig:fixed_ic_sparse_grid}
\end{figure}

\begin{figure}[!htbp]
    \centering
    \begin{minipage}[t]{\textwidth}
        \vspace{0pt}
        \centering
        \subcaptionbox{GO + GC\label{fig:varying_sparse_go_gc}}{\includegraphics[width=0.48\linewidth]{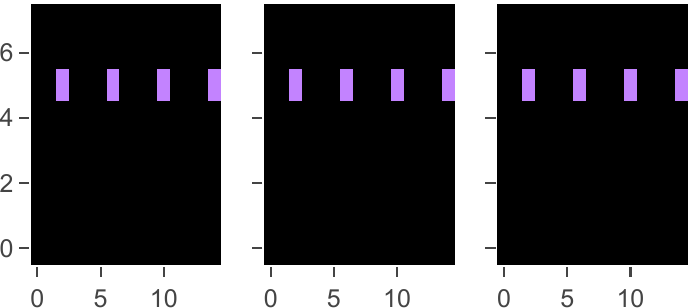}}
        \hfill
        \subcaptionbox{GO + SC\label{fig:varying_sparse_go_sc}}{\includegraphics[width=0.48\linewidth]{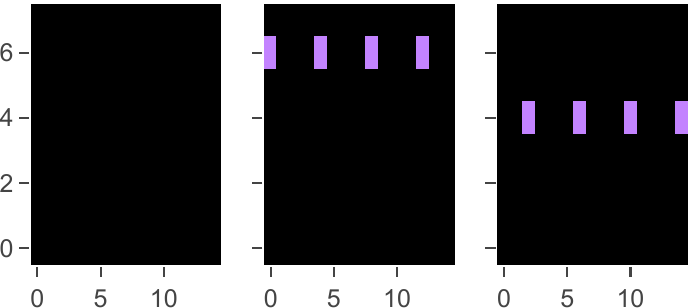}}

        \vspace{0.6em}

        \subcaptionbox{GR + GC\label{fig:varying_sparse_gr_gc}}{\includegraphics[width=0.48\linewidth]{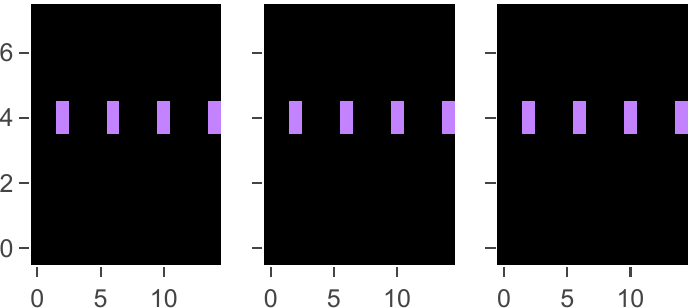}}
        \hfill
        \subcaptionbox{GR + SC\label{fig:varying_sparse_gr_sc}}{\includegraphics[width=0.48\linewidth]{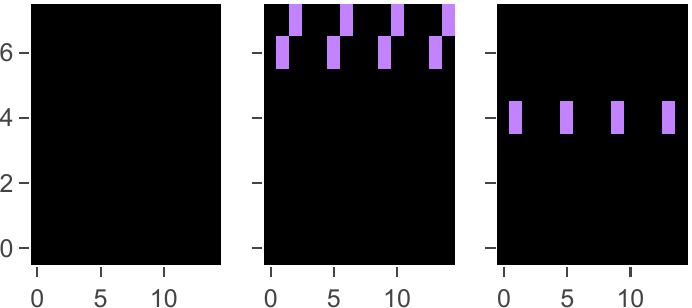}}
    \end{minipage}
    \hfill
    \caption{Varying-IC window sparsity plots for the GO/GR regularizer and channel-grouping combinations. For each combination, three channels are shown in this order: temperature, vertical velocity, and horizontal velocity. The width of each observation plot corresponds to window size 15. Black cells mark pruned inputs, and light-violet cells mark non-pruned inputs.}
    \label{fig:varying_ic_sparse_grid}
\end{figure}

The corresponding GO/GR sensor selections are visualized in Figure~\ref{fig:fixed_ic_sparse_grid} and Figure~\ref{fig:varying_ic_sparse_grid}.
In both settings, the selected rows concentrate on the middle and upper parts of the sensor grid. Separate-channel configurations also tend to deprioritize temperature and retain velocity information, which contrasts with the usual temperature-centric RBC visualization perspective.

\FloatBarrier

\subsection{Learning from Minimal Sensor Sets}\label{sec:minimal_sensor_sets}

As a proof of concept, we also test whether full RL training is possible when each agent observes only a minimal sensor subset. For this experiment, we use the learned mask from the Group Ordered + Grouped Channels setting and construct reduced observations from it. This reduces the observation size per agent from $360$ to $12$.

\begin{figure}[!htbp]
    \centering
    \includegraphics[width=0.8\columnwidth]{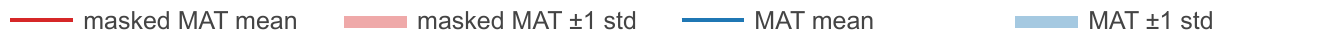}

    \subcaptionbox{}{%
        \includegraphics[width=0.48\textwidth]{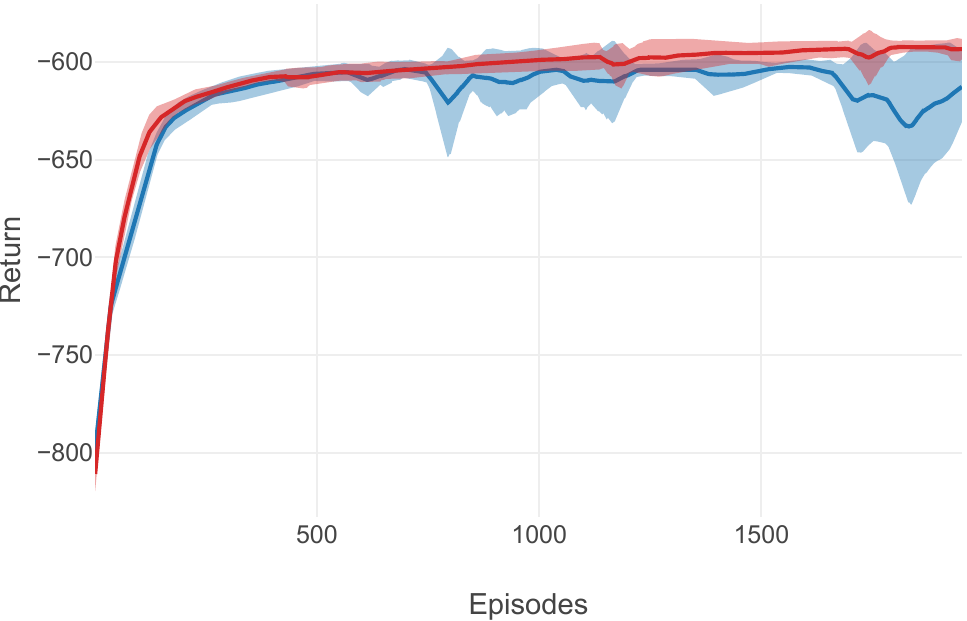}
    }
    \hfill
    \subcaptionbox{}{%
        \includegraphics[width=0.48\textwidth]{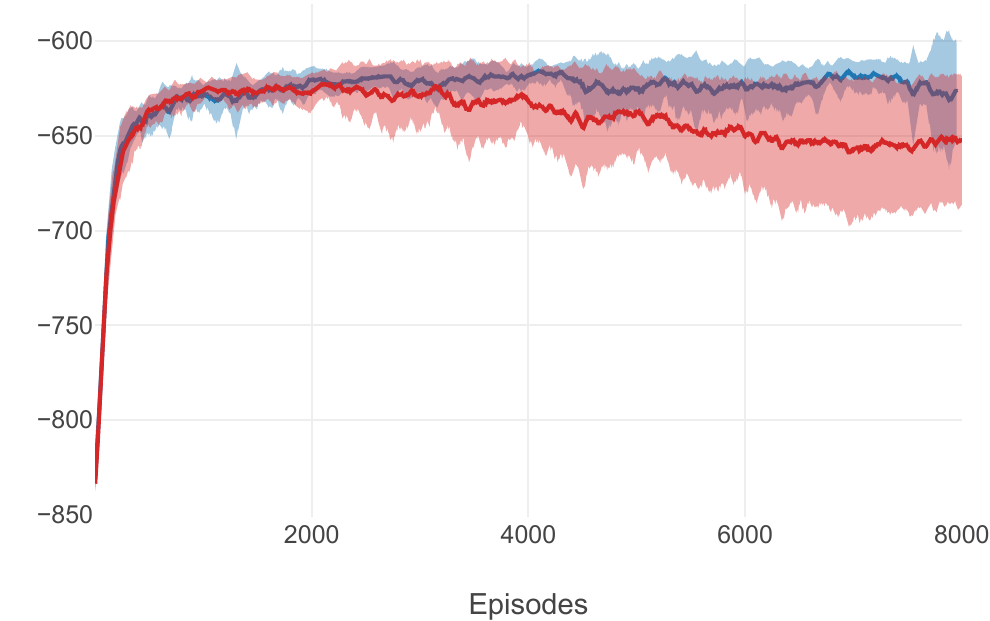}
    }
    \caption{MAT training with full versus minimal sensor sets. Both panels show mean and standard deviation over 10 training runs, with (a) fixed IC and (b) varying IC.}
    \label{fig:minimal_sensor_sets_training}
\end{figure}

It is important to note that a local reward cannot be computed reliably from these minimal observations alone. Therefore, as in the previous experiments, we compute the reward as the negative global instantaneous Nusselt number~(\ref{eq:rbc_nusselt}) evaluated on the full sensor set. Because of this requirement, minimal-sensor training is not directly realizable in reality unless an alternative way to estimate the global instantaneous Nusselt number is available. In simulation, however, this is not a limitation. Figure~\ref{fig:minimal_sensor_sets_training} shows that for fixed IC the minimal-sensor setup reaches the final return regime slightly faster and remains more stable than training on the full sensor set. For varying IC, training with minimal sensors remains feasible but becomes somewhat less stable over longer horizons.
This indicates that extreme sparsity levels like those used here can introduce instabilities in more diverse tasks, so less extreme masks are likely the better choice for training in such settings while still retaining data-throughput savings from reduced observation size.

\FloatBarrier

\section{Conclusion}\label{sec:conclusion}

We presented a sparse-sensing framework for PDE control that combines multi-agent RL, MAT-based expert policies, and supervised apprentice training with grouped regularization. Across fixed and varying initial conditions, MAT showed robust training behavior and the sparse apprentices retained control quality comparable to the dense expert. At the same time, sparsity results were very strong for the proposed grouped methods. In the fixed-IC setting, all GO and GR grouping variants reached maximal sparsity levels in our setup. In the varying-IC setting, GO and GR grouped-channel variants reached maximal sparsity and their separate-channel variants remained near-maximal, with group-ordered outperforming group-reweighted in separate-channel sparsity. The resulting sensor-selection patterns also support the interpretability motivation of this work by indicating which spatial regions and physical state components are most relevant for control.
As a proof of concept, we additionally demonstrated RL training from minimal sensor sets derived from a learned mask, reducing observation size per agent from 360 to 12.

A direct practical implication across applicable setups is that structured sparsification can, in real-world deployments, reduce the required number of probe locations and measured state components and thereby lower sensor hardware and installation effort.
An important next step is to identify real control applications where large windowed sensor evaluations are currently used to compute control-relevant signals.
A second direction is transfer from 2D to 3D RBC. Our observations suggest a targeted hypothesis to test early in that setting, namely starting without temperature measurements and prioritizing velocity components in middle and upper regions.
Because of the much higher dimensionality in 3D, successful transfer of these sparsity patterns could reduce observation bandwidth and inference load.

\section*{Acknowledgements}
Jan Stenner acknowledges support by “ESN4NW: Energieoptimierte Supercomputer-Netzwerke durch die Nutzung von Windenergie” (Grant ID 16ME0617K), funded by the German Federal Ministry of Research, Technology and Space (BMFTR), Germany.

Hans Harder acknowledges support by “SAIL: SustAInable Lifecycle of Intelligent Socio-Technical Systems” (Grant ID NW21-059D), funded by the Ministry of Culture and Science of the State of Northrhine Westphalia (NRW), Germany.

Sebastian Peitz acknowledges funding from the European Research Council (ERC Starting Grant “KoOpeRaDE”) under the European Union’s Horizon 2020 research and innovation programme (Grant agreement No. 101161457).

\section*{Declaration of generative AI and AI-assisted technologies in the manuscript preparation process}

During the preparation of this work, the authors used OpenAI Codex to improve the language and readability of parts of the manuscript. The tool was not used to generate research ideas, perform analyses, create figures, or draw scientific conclusions. After using this tool, the authors reviewed and edited the content as needed and take full responsibility for the content of the published article.

\clearpage

\bibliographystyle{unsrtnat}
\bibliography{references}

\clearpage
\appendix
\section{General Group Construction and Overlap Proof}\label{app:grouping}

\subsection{General Construction of Grouping Classes}\label{app:group-construction}

To formalize the grouping process in a general setting, let $t_o$ be an $N$-th order observation tensor of sensor measurements, whose flattened form $o$ is used as model input. Let $D=\{1,2,\ldots,N\}$ be its set of modes, let $d_k$, $k\in D$, denote dimensions, and define
\begin{align*}
    \mathcal{I}_o = \prod_{k\in D}[\,d_k\,] = [d_1]\times[d_2]\times\cdots\times [d_N],\qquad
    n_o = \prod_{k\in D} d_k,
\end{align*}
so that $o=\mathrm{vec}(t_o)\in\mathbb R^{n_o}$. In this appendix section, $\mathcal I_o$ is the local coordinate system of one window. To separate local from global coordinates, let $\mathcal G\subseteq\mathbb Z^N$ be the global sensor grid and $\mathcal C\subseteq\mathcal G$ the admissible agent centers.

Assume $D_o \subseteq D$ is a non-empty set of overlap modes and that $t_o$ contains a local center index $i_c$. Let $\delta_i \in \mathbb{Z}_{>0}$ denote center spacings in each mode $i\in D_o$. Define
\begin{align*}
    \mathcal E_o=\{\eta\in\mathbb Z^N:\ i_c+\eta\in\mathcal I_o\},\qquad
    \phi:\{1,\dots,n_o\}\to\mathcal E_o.
\end{align*}
Here, $\phi$ is the flattening bijection that maps flattened indices of $o$ to local offsets. Hence, $i_c+\phi(i)$ are the local window coordinates associated with index $i$.
For $c\in\mathcal C$ and $i\in\{1,\dots,n_o\}$, define the global-point map
\begin{align*}
g(c,i)=c+\phi(i).
\end{align*}
This map returns the global grid point addressed by local index $i$ in the window centered at $c$.
For example, with $N=2$ and row-wise flattening,
\begin{align*}
t_o \in \mathbb R^{n\times m},
\quad &
o = \mathrm{vec}(t_o)\in\mathbb R^{nm},\\
\phi(i) = \eta=(q_1-(i_c)_1,\;q_2-(i_c)_2)
\quad\Longleftrightarrow\quad &
i = (q_1-1)\,m + q_2, \quad \forall (q_1,q_2)\in \mathcal{I}_o=[n]\times[m].
\end{align*}

Define anchors and stencils by
\begin{align*}
    \mathcal{W} = \{w \in \mathbb{Z}^N:\quad & w_i = \delta_i\,n_i,\ n_i\in\mathbb Z,\ \forall i\in D_o,\\
    & w_j=0 \quad \forall j\in D\setminus D_o\},\\
    \mathcal{H} = \{h \in \mathbb{Z}^N:\quad &  0 \le h_i \le \delta_i-1 \quad \forall i\in D_o,\\
    &  -(i_c)_j \le h_j \le d_j-(i_c)_j \quad \forall j\in D \setminus D_o\},
\end{align*}
and the relevant anchors
\begin{align*}
    \mathcal W_{\mathrm{rel}}=\{w\in\mathcal W:(i_c+w+\mathcal H)\cap\mathcal I_o\neq\varnothing\}.
\end{align*}
Assume a reference center exists:
\begin{align}
\exists\,c_\star\in\mathcal C\quad\text{s.t.}\quad c_\star+w\in\mathcal C\ \ \forall w\in\mathcal W_{\mathrm{rel}}.
\label{ass:refcenter}
\end{align}

For any distinct $w,v\in\mathcal W_{\mathrm{rel}}$, the finite sets
\begin{align*}
  (i_c+w + \mathcal H)\,\cap\,\mathcal I_o
  \quad\text{and}\quad
  (i_c+v + \mathcal H)\,\cap\,\mathcal I_o
\end{align*}
are disjoint.

For every $h\in\mathcal H$, define
\begin{align*}
    \mathcal{I}_h = \bigl\{ \phi^{-1}(w + h) : w\in\mathcal W_{\mathrm{rel}}, \; i_c + w + h \in \mathcal{I}_o \bigr\} \; \subseteq \{ 1,\dots,n_o \}.
\end{align*}
These sets are pairwise disjoint and induce the equivalence relation
\begin{align*}
i \sim j \quad\Longleftrightarrow\quad \exists\,h\in\mathcal H:\; i,j \in \mathcal I_h.
\end{align*}
Under (\ref{ass:refcenter}), the following characterization holds (proof in Appendix~\ref{app:overlap-proof}): two local indices, viewed from two different window centers, point to a common global grid point if and only if they lie in a common set $\mathcal I_h$. Equivalently, this overlap can be written in operational form as
\begin{align*}
    c_i + v + h = c_j + w + h,\qquad v,w\in\mathcal W_{\mathrm{rel}},\ h\in\mathcal H,
\end{align*}
for suitable centers $c_i,c_j\in\mathcal C$.

\subsection{Input-Channel Grouping Extension}\label{app:channel-grouping}

For channel-coupled grouping, let $m$ denote the channel mode and let $D$ be the set of all non-channel modes of $t_o$. Let $\mathcal E_o$ be the corresponding local offset set in these modes. Then $\phi$ is a surjective map from flattened indices of $o$ to offsets in $\mathcal E_o$. Let $d_m$ be the channel-mode dimension and let $\phi^{-1}_1,\dots,\phi^{-1}_{d_m}$ be channel-specific inverse maps back to flattened indices. For each $h\in\mathcal H$, define
\begin{align*}
    \mathcal{I}_{h;i} &= \bigl\{ \phi^{-1}_i(w + h) : w\in\mathcal{W}_{\mathrm{rel}}, \; i_c + w + h \in \mathcal{I}_o \bigr\} \; \subseteq \{ 1,\dots,n_o \},\\
    \mathcal{I}_h &= \bigcup_{i=1}^{d_m} \mathcal{I}_{h;i}.
\end{align*}
This couples pruning decisions across channels at the same spatial locations.

\subsection{Proof of the Overlap Characterization}\label{app:overlap-proof}

\begin{figure}[t]
    \centering
    \includegraphics[width=0.8\columnwidth]{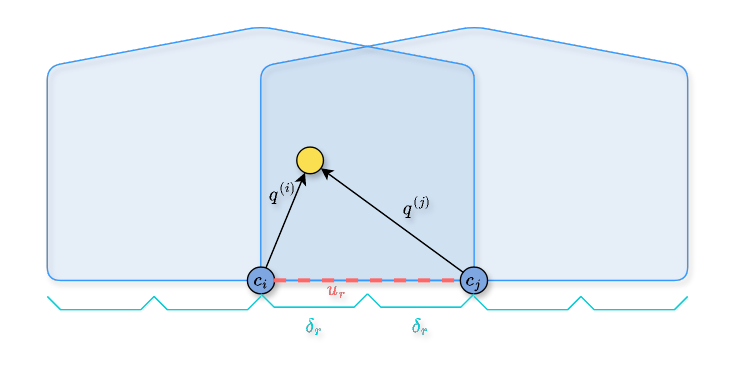}
    \caption{Illustration of the overlap construction in two modes. One mode is an overlap mode, where the centers $c_i$ and $c_j$ are separated by $2\delta_r$. Offsets $q^{(i)}$ and $q^{(j)}$ define the global points $c_i+q^{(i)}$ and $c_j+q^{(j)}$, which coincide in the shown example. The outer boxes indicate the corresponding local window size around each center.}
    \label{fig:overlap_proof}
\end{figure}

\textbf{Proposition.}
For $i,j\in\{1,\dots,n_o\}$, the following are equivalent:
\begin{align*}
\text{(A)}\quad &\exists\,c_i,c_j\in\mathcal C:\ g(c_i,i)=g(c_j,j),\\
\text{(B)}\quad &\exists\,h\in\mathcal H:\ i,j\in\mathcal I_h.
\end{align*}

\textit{Proof.}
\emph{(A)$\Rightarrow$(B):} Let (A) hold and set $q^{(i)}:=\phi(i)$, $q^{(j)}:=\phi(j)$. Then
\begin{align*}
c_i+q^{(i)}=c_j+q^{(j)}
\quad\Longrightarrow\quad
q^{(i)}-q^{(j)}=c_j-c_i=:u.
\end{align*}
By anchor compatibility, $u\in\mathcal W$. Hence, for each overlap mode $r\in D_o$, there exists $k_r\in\mathbb Z$ such that
\begin{align*}
q^{(i)}_r = q^{(j)}_r + u_r = q^{(j)}_r + k_r\delta_r.
\end{align*}
Figure~\ref{fig:overlap_proof} illustrates this relation for a representative two-dimensional case.
Therefore
\begin{align*}
\mathrm{mod}(q^{(i)}_r,\delta_r)=\mathrm{mod}(q^{(j)}_r,\delta_r), \qquad r\in D_o,
\end{align*}
where $\mathrm{mod}(\cdot,\delta_r)$ denotes the nonnegative remainder in $\{0,\dots,\delta_r-1\}$.
For non-anchor modes $j\in D\setminus D_o$, we have $u_j=0$ by definition of $\mathcal W$, hence
\begin{align*}
q^{(i)}_j=q^{(j)}_j.
\end{align*}

To construct anchors in $\mathcal W$, define $h\in\mathcal H$ coordinate-wise by
\begin{align*}
h_r&:=\mathrm{mod}\!\bigl(q^{(i)}_r,\delta_r\bigr)=\mathrm{mod}\!\bigl(q^{(j)}_r,\delta_r\bigr), && r\in D_o,\\
h_j&:=q^{(i)}_j=q^{(j)}_j, && j\in D\setminus D_o.
\end{align*}
Now set
\begin{align*}
v:=q^{(i)}-h,\qquad w:=q^{(j)}-h.
\end{align*}
Then $q^{(i)}=v+h$ and $q^{(j)}=w+h$. Moreover, for each $r\in D_o$,
\begin{align*}
v_r = q^{(i)}_r-h_r\in \delta_r\mathbb Z,\qquad
w_r = q^{(j)}_r-h_r\in \delta_r\mathbb Z,
\end{align*}
and for each $j\in D\setminus D_o$,
\begin{align*}
v_j=q^{(i)}_j-h_j=0,\qquad w_j=q^{(j)}_j-h_j=0.
\end{align*}
so $v,w\in\mathcal W$. Since $q^{(i)},q^{(j)}\in\mathcal E_o$ and $q^{(i)}=v+h$, $q^{(j)}=w+h$, we also have $v,w\in\mathcal W_{\mathrm{rel}}$. Therefore
\(
i=\phi^{-1}(v+h)\in\mathcal I_h
\)
and
\(
j=\phi^{-1}(w+h)\in\mathcal I_h
\),
so (B) holds. In particular, we have shown the concrete representation
\begin{align*}
\phi(i)=q^{(i)}=v+h,\qquad \phi(j)=q^{(j)}=w+h,\qquad c_i+v+h=c_j+w+h.
\end{align*}

\emph{(B)$\Rightarrow$(A):} Let (B) hold. Then there exists $h\in\mathcal H$ and anchors $v,w\in\mathcal W_{\mathrm{rel}}$ such that
\(
i=\phi^{-1}(v+h),\ j=\phi^{-1}(w+h)
\),
equivalently
\(
\phi(i)=v+h,\ \phi(j)=w+h
\).
By (\ref{ass:refcenter}), choose
\(
c_i:=c_\star+w,\ c_j:=c_\star+v
\),
which are both in $\mathcal C$. Then
\begin{align*}
g(c_i,i)&=c_i+\phi(i)=(c_\star+w)+(v+h)=c_\star+v+w+h,\\
g(c_j,j)&=c_j+\phi(j)=(c_\star+v)+(w+h)=c_\star+v+w+h.
\end{align*}
So $g(c_i,i)=g(c_j,j)$, and thus (A) holds.

Therefore (A) and (B) are equivalent. \hfill$\square$

\end{document}